\documentstyle[a4wide,12pt,epsfig]{article}
\newcommand{\be}{\begin{equation}}
\newcommand{\ee}{\end{equation}}

\begin{document}
\begin{flushright}
 hep-ph/9605254\\
IISc-CTS-10/96\\
LNF-96/019(P)
\end{flushright}
\begin{center}
\begin{Large}
Mini-jet Total Cross-sections and Overlap Functions through 
Bloch-Nordsieck Summation\footnote{Presented by G. Pancheri at {\it 
Hadron Structure '96}, Stara Lesna, Feb. 12-17, 1996}
\end{Large}
\vspace{1 cm}

A. Corsetti\\
INFN, Physics Department, University of Rome La Sapienza, Rome, Italy\\

R. M. Godbole\footnote{On leave of absence from Dept. of Physics, 
Uni. of Bombay, Bombay, India}
\\Center for Theoretical Studies, Indian Institute of Sciences\\
Bangalore 560 012, India\\

A. Grau\\Departamento de F\'\i sica Te\'orica y del Cosmos,
Universidad de Granada, Spain\\

G. Pancheri \\
 INFN - Laboratori Nazionali di Frascati , I00044 Frascati\\
 
 and\\
 
 Y.N. Srivastava\\
Physics Department and INFN, University of Perugia, Perugia, Italy
\end{center}
\vspace{0.5cm}
\begin{abstract}
Predictions for total inelastic cross-sections for photon induced
processes are discussed in the context of the QCD-inspired minijet model.
 Large theoretical uncertainties exist, some of them related to
 the parton distributions of hadrons in impact parameter space. A
 model for such distribution is presented,  based on 
 soft gluon summation. This model 
incorporates (the salient features of
distributions obtained from) the intrinsic transverse momentum
behaviour of hadrons. Under the assumption that the intrinsic behaviour is
dominated by soft gluon emission  stimulated by the scattering process,
 the b-spectrum becomes softer and softer as the
 scattering energy increases. In minijet models for the inclusive 
  cross-sections, 
 this will counter the increase from $\sigma_{jet}$ .
\end{abstract}

The impact of parton scattering  on the rise 
of inclusive cross-sections with energy  
was suggested  by Cline and Halzen \cite{CLINE}, after such rise was 
first observed in proton-proton
collisions at ISR .  Minijet models were put forward to describe
quantitatively the further rise at higher energy  
\cite{CR,GAISSER,PS} and eikonal minijet models \cite{DURAND,BLOCK,TRELEANI,
CAPELLA} 
were subsequently developed to include 
 an increasing number of partonic collisions in QCD
resulting from the rapid rise in gluon densities.  
Recent measurements of photo- and hadro-production total
 cross-sections
\cite{HERA,TEVA} in 
energy regions where QCD processes dominate, confirmed the rising trend and 
have been 
confronted  with
theoretical predictions obtaining varying degrees of success \cite{GODBOLE}.

The simplest mini-jet model \cite{GAISSER} was written as
\be
\label{miniHalzen}
\sigma_{inel}(s)=\sigma_{soft}+\sigma_{jet}(p_{tmin},s)
\ee
with the rise controlled only by  an energy dependent parameter,
namely $p_{tmin}$, which regulated the Rutherford scattering divergence in
the QCD jet-cross-section. The lack of unitarity of this model was amended
in the  eikonalized mini-jet model, where  
 inelastic  hadron-hadron cross-sections are written
as
\be
\label{eiksigma}
\sigma_{inel}=\int d^2{\vec b} \big[ 1-e^{-n(b,s)}\big].
\ee
Here the average number of collisions at impact parameter b is given by
\be
\label{nb}
 n(b,s)=A(b)\sigma (s)
\ee
with 
fixed $p_{tmin}$. In this version,  the   excessive QCD rise,
 controlled by $p_{tmin}(s)$ in the non-unitarized
models, was  softened through eikonalization and introduction of
the overlap function $A(b)$, which thus became
 a key ingredient of all models with a QCD component. 
This function, which  describes 
matter  distribution in  impact parameter space, in most applications
\cite{DURAND}
 has been assumed to be
the Fourier transform of the product of  hadronic form factors of the colliding 
particles. 
In other models \cite{PYTHIA}, a gaussian shape has been preferred, 
thus relating $A(b)$ to the 
intrinsic transverse momentum distribution of  partons in the colliding
hadrons. In either case, 
detailed information on $A(b)$ relies on parameters
to be determined  case by case. However, while 
direct measurements of the EM form factors are  available for nucleons 
and pseudoscalar mesons, experimental information regarding photons or
other 
hadrons such as vector mesons is lacking. The same observation applies
to the intrinsic-$p_t$ interpretation for the spatial distribution of partons 
in vector mesons or photons. Thus the extension of this model to
photonic cross-sections unveils  one of its main drawbacks, viz.
lack of a fundamental description 
of
parton b-distributions.
To reduce the uncertainties in the QCD 
description of the
rise of the inelastic cross-section, and allow this model to graduate
to QCD respectability, it is mandatory to arrive
 at a QCD description of the overlap
function $A(b)$.

 For the case of photonic processes, there are further 
uncertainties related to the hadronic behaviour of photons.
The model has been  adapted to photonic processes by writing
the inelastic cross-section as
 
\begin{equation}
\label{eikonal}
\sigma^{inel}_{ab} = P^{had}_{ab}\int d^2\vec{b}[1-e^{-n(b,s)}]
\end{equation}
where  $P^{had}_{ab}$  gives the probability that both colliding particles
$a,b$ be in a hadronic state\cite{LADCOLL} and  
$n(b,s)=n_{soft}(b,s)+n_{hard}(b,s)$. 
Here  $n_{soft}(b,s)$ contains the non-perturbative part of the 
cross-section from 
which the factor  $P_{ab}^{had} $ has already been factored out and 
 the hard, QCD 
contribution to the average number of collisions at a given impact
parameter $\vec{b}$ is given by
\begin{equation}
\label{}
n_{hard}(b,s)=A_{ab} (b){{1}\over{P^{had}_{ab}}}
\sigma^{jet}_{ab}
\end{equation} 
$\sigma^{jet}_{ab} $ is the hard part of the cross--section.  We 
then have
$$
P_{\gamma p}^{had} = P_{\gamma}^{had} ; 
P_{\gamma \gamma}^{had} =  (P_{\gamma}^{had})^2 ; \ \ \ P_{pp}^{had}=1.
$$

The   predictions of the eikonalised mini-jet model  
\cite{minijets} for photoproduction processes therefore
depend on 1) the assumption of one or more eikonals 2)  the hard jet 
cross-section 
$\sigma^{jet}_{ab}=\int_{p_{tmin}} {{d^2\hat{\sigma}}\over{dp_t^2}} dp_t^2$ 
which in turn depends on the minimum 
$p_t$ above which one can expect perturbative QCD to hold, viz. $ p_{tmin}$,
and the  parton densities in the colliding particles $a$ and $b$, 
3) the soft cross--section $\sigma^{soft}_{ab}$ 4) the overlap function
$ A_{ab}(b) $,  usually written as 
\begin{equation}
\label{aob}
A_{ab}(b)={{1}\over{(2\pi)^2}}\int d^2\vec{q}{\bf F}_a(q) {\bf F}_b(q) 
e^{i\vec{q}\cdot \vec{b}}
\end{equation}
 where ${\bf F}$ is the Fourier transform of the b-distribution
of partons in the colliding particles and 5) last, but not the least,
$P_{ab}^{had}$.

To study the parameter dependence of this model,
one can restrict   attention to  a single eikonal : more
eikonal terms although improving the fits, de facto introduce 
new sets of parameters and very much reduce the predictivity of the
model. 
The hard jet cross-sections have  been evaluated  in LO perturbative QCD using 
two different photonic parton densities  DG \cite{DG} and GRV \cite{GRV}. The
dependence of $\sigma_{ab}^{jet}$ on $ p_{tmin}$ for DG densities is given in 
Ref. \cite{GODBOLE}. Clearly this dependence is strongly correlated with
the parton densities used. Here we shall only show the results for eikonalised
mini--jet  cross-sections using GRV densities. Further, for the purposes of 
this note, we try to  estimate $\sigma^{\gamma \gamma}_{soft}$ 
from $\sigma^{\gamma p}_{soft}$ which in turn is  determined by a 
fit to the photoproduction data. For $\gamma \gamma$ collisions, we use the 
Quark Parton Model  suggestion  $\sigma_{soft}^{\gamma \gamma} = {{2}\over{3}} 
\sigma_{soft}^{\gamma p}$. 

In the original use of the eikonal model, the overlap function $A_{ab} (b)$ of
eq. \ref{aob}  was obtained using for ${\bf F}$ the electromagnetic form 
factors and  thus, for
 photons, 
 a  number of authors 
\cite{SARC, FLETCHER} have assumed 
for ${\bf F }$    the pion pole expression,  on the basis of Vector
 Meson Dominance (VMD).
As mentioned, another possibility is that the b-space 
distribution of partons in the photon is the Fourier transform 
of their intrinsic 
transverse momentum distributions. 
For protons this has 
been assumed to correspond to a gaussian
shape.  For photons, the perturbative part
 \cite{pertint}
of the intrinsic transverse momentum  
has been
suggested to correspond to the functional expression 
\begin{equation}
\label{intrinsicphot}
{{d N_{\gamma}}\over{dk_t^2}}={{1}\over{k_t^2+k_o^2}}
\end{equation}
\thispagestyle{empty}
Recently this expression was verified by the ZEUS \cite{ZEUS} Collaboration,
with $k_o=0.66 \pm 0.22 \ GeV$. It is interesting 
to notice that
 for photonic collisions
the overlap function will have the same analytic expression 
for both  ans\"atze for {\cal F}: the VMD inspired pion form factor or the intrinsic
transverse momentum; the only difference being that the former corresponds to
a fixed value of $k_0 = 0.735 \;\; GeV$ whereas  the latter allows  to 
vary the 
value of the parameter $k_0$.
  Thus both 
 possibilities can be easily studied by simply changing $k_0$ appropriately.
 The overlap function, 
which for proton-proton collisions would be given by
\begin{equation}
\label{aobpp}
A^{FF}_{p p}(b)=\int {{d^2{\vec Q}\over{(2\pi)^2}}} e^{i {\vec Q}\cdot {\vec b}} 
\big({{\nu^2}\over{Q^2+\nu^2}}\big)^4={{b \nu^2  \sqrt{\nu^2}}
\over{96 \pi}} {\cal K}_{3}(b \sqrt{\nu^2})\ \ \ \ \ \ \ \nu^2=0.71\ GeV^2
\ee
as proposed   by L.Durand et al. \cite{DURAND}
in the first eikonal mini-jet model for proton-proton collisions, for
$\gamma p$ collision would become
\begin{equation}
\label{aobgp}
A_{\gamma p}^{FF}={{1}\over{4 \pi}}{{\nu^2 k_o^2}\over{k_o^2-\nu^2}}
\Big[
\nu bK_1(\nu b)-{{2\nu^2}\over{k_o^2-\nu^2}}
\left[ K_0(\nu b)-K_0(k_o b)\right]\Big]
\end{equation}
and for $\gamma \gamma$ collisions
\begin{equation}
\label{aobgg}
A_{\gamma \gamma}^{FF}(b)={{1}\over{4 \pi}} k_o^3 b K_1(b k_o)
\end{equation}

As for $P^{had}_\gamma$, this is clearly expected to be ${\cal O} (\alpha_{em})$
and from  VMD one would expect $1/250$.  It should 
be noticed that 
the eikonalised minijet cross--sections do not depend  on
$A_{\gamma \gamma}$ and $P^{had}_{\gamma \gamma}$ separately, 
but depend only on the ratio of the two \cite{drees1,dgo10};
which for our ans\"atze for $A_{\gamma \gamma}$ means ratio of $k_0$ and 
$P^{had}_{\gamma \gamma}$. From
phenomenological considerations
\cite{FLETCHER}and fits to HERA data,  fixing $k_o=0.735\ GeV$ 
 one finds a value $P^{\gamma}_{had}\approx 1/200$, which 
indicates at these energies a non-VMD component of $\approx 20\%$.

As mentioned, 
the QCD description requires the definition of 
$p_{tmin}$. From HERA data, one  notices that while lower values of $p_{tmin}$,
i.e. 1.4 GeV, can
be invoked to describe the beginning of the rise, a higher value, i.e. 2.0 or 
even 2.5 GeV, is better suited to describe the rise at higher energy.
 In Fig.1a
we show the fit to HERA data obtained with the above parameters, using
a purely phenomenological fit to determine the
non-perturbative part of the cross-section.

Having thus established the range of variability of the quantities 
involved in the calculation of
 total photonic cross sections, we now proceed to calculate and compare with
existing data the eikonalized minijet cross-section for
$\gamma \gamma$ collisions. 
For
photon photon collisions,  we use the central value $p_{tmin}=2.0 \ GeV$.
We also use  $P_{\gamma}^{had} =1/204$,
and A(b) from eq.(\ref{aobgg}) with 3 different values 
of $k_o$ which correspond to  values within two  standard deviations
 from the ZEUS \cite{ZEUS}
collaboration value.
  \begin{figure}[ht]
\begin{center}
\leavevmode
\mbox{\epsfig{file=starelesnafig1.ps,width=0.7\textwidth,bbllx=30pt,bblly=70pt,bburx=570pt,bbury=750pt,angle=90}}
\end{center}
\caption{Total inelastic photon-proton and photon-photon cross-section 
as described in the text.}
\label{gamgam}
\end{figure}
Our predictions for $\gamma \gamma$ collisions are shown in Fig. (1b). A 
comparison
with existing  data shows that data points are better fitted by a higher
larger value of $k_0$, and we choose $k_0=1\ GeV$. 

As stressed,the theoretical description is rather unsatisfactory and we now
move to present a model for the overlap function which,in principle, should
 allow for a clearer predictability
\thispagestyle{plain}
and  to provide an expression   for $A(b)$ which
could be applied to various cases of interest.  We shall use  
Bloch-Nordsieck techniques to sum soft gluon transverse momentum distributions
 to all orders and compare 
our results with both the intrinsic transverse momentum approach as well as
the form factor approach. In what follows, we shall first illustrate
the Bloch-Nordsieck result and show that  it gives 
a gaussian fall-off with an intrinsic transverse size consistent with
MonteCarlo models \cite{PYTHIA}.
We then calculate the 
relevant distributions and discuss their phenomenological
application.

In ref.(\cite{NOSTRO}) it has been proposed  that in hadron-hadron collisions,
 the b-distribution of 
partons in the colliding hadrons is the Fourier transform of the 
transverse momentum distribution resulting from
soft gluon radiation emitted by quarks as the hadron 
breaks up because
of the collision. This distribution is obtained by summing soft 
gluons to all orders,
with a technique amply discussed in the literature \cite{EPT,GPSS}. 
The resulting
expression\cite{DDT,PP} is
\thispagestyle{plain}
\begin{equation}
\label{BNPT}
{\cal F_{BN}}(K_\perp)={{1}\over{2 \pi}}\int b db J_0(b K_\perp) e^{-h(b;M,
\Lambda)}
\end{equation}
with 
\begin{equation}
\label{hdb}
h(b;M,\Lambda)={{2 c_F}\over{\pi}}\int_0^M {{dk_\perp}\over{k_\perp}}\alpha_s({{k^2_\perp}
\over{\Lambda^2}})\ln{{M+\sqrt{M^2-k_\perp^2}}\over{M-\sqrt{M^2-k_\perp^2}}}
[1-J_0(k_\perp b)]
\end{equation}
where $c_F=4/3$ and the  hadronic scale M accounts for the maximum energy 
allowed in
a single ($k^2=0$) gluon emission.

\thispagestyle{plain}
 The definition given in eqs.(\ref{eiksigma},\ref{nb}), requires for its 
consistency a  normalized  b-distribution, i.e.  
\be
\label{norm} 
\int d^2 {\vec b} A(b) = 1.
\ee 
so that the  proposed Bloch-Nordsieck expression for the overlap function
 $A(b)$, satisfying the above normalization, reads
\be
\label{ourAB}
A_{BN}= {{e^{-h(b;M,\Lambda)}}\over
{2\pi \int bdb 
 e^{-h(b;M,\Lambda)}}}
\end{equation}
An inspection of eq.(\ref{hdb}), immediately 
poses the problem of extending the known asymptotic freedom expression 
for $\alpha_s$ to the very small $k_\perp$ region. 
To avoid the small $k_\perp$ divergence in eq.(\ref{hdb}), 
it has been customary to   introduce a lower cut-off in $k_\perp$ 
and freeze $\alpha_s$ at $k_\perp=0$, i.e. to put
\begin{equation}
\label{altarelli}
\alpha_s(k_\perp^2)={{12 \pi}\over{33-2 N_f}} {{1}\over{\ln[(k_\perp^2+a^2 
\Lambda^2)/\Lambda^2)]}}
\end{equation}
with $a=2$ in ref. \cite{ALTARELLI}. For applications where the
scale $M$ is large  (e.g., W-transverse momentum distribution calculations)
eq.(\ref{hdb}) is dominated by the (asymptotic) logarithmic behaviour and 
the small $k_\perp$-limit, albeit theoretically crucial, is not very relevant 
phenomenologically. However, this is not case in the present context,
where we are dealing with soft gluon emission in low-$p_t$ physics 
(responsible for large cross-sections). The typical scale 
of such peripheral interactions,  is that of the hadronic masses, i.e.
we expect $M\sim {\cal O}(1\div2 \ GeV)$ and the small $k_\perp$ limit
plays a basic role. This can be  appreciated  on a qualitative basis, by 
considering the limit  $ b M<<1$ of eq. (\ref{hdb}).
  In this region, one can approximate 
$1-J_0(kb)\approx b^2k^2/4$, to obtain
\begin{equation}
\label{intkdk}
h(b;M,\Lambda)\approx b^2\ A
\ee
with 
\be
\label{A}
 A= {{c_F}\over{4\pi}}
\int  dk^2 \alpha_s({{k^2}\over{\Lambda^2}}) \ln
{{4 M^2}\over{k^2}}
\end{equation}
One obtains a function $h(b;M,\Lambda)$ 
with  a gaussian fall-off as in models where 
  $A(b)$ is the Fourier transform of an intrinsic transverse momentum 
  distribution of partons, i.e.  
  \newline $\exp(-k_\perp^2/4 A^2)$. Note that the relevance of an integral
 similar
to the one in eq.(\ref{A}) has been recently discussed in connection to
 hadronic event shapes \cite{DW}.

\thispagestyle{plain}
 Our choice for the infrared behaviour of $\alpha_s$ for a
  quantitative description of the  distribution in eq.(\ref{hdb}),
  does not follow eq.(\ref{altarelli}), but  is inspired by the
   Richardson potential
  for quarkonium bound states \cite{RICHARDSON}. In 
 a number of related
applications \cite{INTRDY,ourWBOSON}, it has been  proposed to calculate 
the above integral using the  following expression for $\alpha_s$ :
\begin{equation}
\label{alphaRich}
\alpha_s(k_\perp)={{12 \pi }\over{(33-2N_f)}}{{p}\over{\ln[1+p({{k_\perp}
\over{\Lambda}})^{2p}]}}
\end{equation}
which coincides with the usual one-loop expression for large (relative
to $\Lambda$) values of $k_\perp$, while going to a singular
 limit for small $k_\perp$. For the special case $p=1$ such an
$\alpha_s$ coincides with one used in the 
 Richardson potential \cite{RICHARDSON},
and which  incorporates - in a compact expression - 
 the high-momentum limit demanded by asymptotic freedom as well as 
 linear quark confinement in the static limit. In \cite{INTRDY} we have
generalized Richardson's ansatz to values of $p\le 1$. For $1/2< p\le 1$,
this corresponds to a confining potential rising less than linearly
with the interquark distance $r $. 
The range $p\neq 1$ has an important advantage, 
i.e., it allows the integration in eq.(\ref{hdb}) to converge for all 
values of $k_\perp$.
\thispagestyle{plain}
For the motivations given in \cite{INTRDY}
the value $p=5/6$ was chosen  in previous calculations of  
the transverse momentum distribution of Drell Yan pairs \cite{INTRDY,NAKW}.

Having set up our formalism, we shall now examine  its implications. 
The distribution $A(b)$ depends upon the hadronic
scale M in the function $h(b)$. This scale
   depends upon the energy of
the specific subprocess
and, through this, upon  the hadron scattering energy.
It  plays a crucial 
role, just as it did for the Drell-Yan  process, where the 
 expression of eq.(\ref{BNPT}) has been successfully 
\cite{GRECO,HALZEN,NAKW,ALTARELLI} used to describe the transverse momentum  
distribution of the
time-like virtual photon  or W-boson. In these cases\cite{GRECO,NAKW},
the scale M was found to be energy
dependent and to vary between $\sqrt{Q^2}/4$ and $\sqrt{Q^2}/2$. 
In the
calculation of the transverse momentum distribution of a lepton
pair produced in  quark-antiquark
annihilation \cite{GRECO},
the scale $M$ was obtained as
  the maximum transverse momentum allowed by
kinematics to a single gluon emitted by the initial $q{\bar q}$ pair of
 c.m.energy 
$\sqrt{{\hat s}}$ in the process
\be
\label{processDY}
q {\bar q} \rightarrow g + \gamma(Q^2)
\ee

In the Drell-Yan case, one needed $h(b)$ to calculate
 the transverse momentum distribution
of the  lepton pair, here we use it to evaluate the average number of
partons in the overlap region  of two colliding hadrons. In this case 
$e^{-h(b)}$ is the {\cal F}-transform of the transverse momentum distribution
induced by initial state radiation  in the process 
\be
\label{process}
q{\bar q} \rightarrow  \ jet \ jet  + X
\ee
where the jet pair   in process 
(\ref{process}) is the one produced through gluon-gluon or other 
parton-parton scattering with total jet-cross-section $\sigma_{jet}$ and 
X can also  include the quark-antiquark pair which continues
undetected after  emission of the  gluon pair   which stimulated 
the initial state bremsstrahlung.
  We work in a 
no-recoil approximation, where the transverse momentum of
 the jet pair is balanced by the emitted soft gluons.
 Then the maximum transverse momentum allowed to a single gluon
is given by  
\be
\label{qmaxjets}
q_{max}({\hat s})={{ \sqrt{{\hat s}} }\over{2}}
 (1-{{{\hat s_{jet}}}\over{{\hat s}}})
\ee
with $\sqrt{{\hat s_{jet}}}$ being the jet-jet invariant mass 
 over which
one needs to perform  further integrations. 
\thispagestyle{plain}
An improved eq.(\ref{nb}) now reads
\be
\label{nbimp}
n(b,s)=n_{soft}(b,s)+\sum_{i,j,}\int{{dx_1}\over{x_1}}
\int{{dx_2}\over{x_2}} f_i(x_1)f_j(x_2)\int dz \int dp_t^2
A_{BN}(b,q_{max}){{d\sigma}\over{dp_t^2 dz}}
\ee
where $f_i$ are the quark densities
 in the colliding hadrons, 
  $z={\hat s_{jet}}/(sx_1x_2)$,  and ${{d\sigma}\over{dp_t^2 dz}}$ is the
  differential cross-section for process (\ref{process})
   for a given $p_t$ of the
  produced jets. \thispagestyle{plain}
  
Unlike the usual expressions for $n(b,s)$, eq.(\ref{nbimp}) does not
exhibit factorization between the longitudinal and transverse
degrees of freedom since the distribution $A_{BN}$ depends upon the
quark subenergies. Factorization can be obtained however, through an
averaging process whereupon 
one     can factorize 
  the b-distribution in eq.(\ref{nbimp}), 
  by evaluating $A_{BN}$ with $q_{max}$ at its mean value, i.e. write
  \be
\label{nbimpmean}
n(b,s)=n_{soft}(b,s)+A_{BN}(b,<q_{max}(s)>)\sigma_{jet}
\ee
with
\be
\sigma_{jet}=\sum_{i,j,}\int{{dx_1}\over{x_1}}
\int{{dx_2}\over{x_2}} f_i(x_1)f_j(x_2)\int dz \int dp_t^2
{{d\sigma}\over{dz dp_t^2}}
\ee
and
\be
\label{qmaxav}
<q_{max}(s)>={{\sqrt{s}} 
\over{2}}{{ \sum_{i,j}\int {{dx_1}\over{ x_1}}
f_{i/a}(x_1)\int {{dx_2}\over{x_2}}f_{j/b}(x_2)\sqrt{x_1x_2} \int dz (1 - z)}
\over{\sum_{i,j}\int {dx_1\over x_1}
f_{i/a}(x_1)\int {{dx_2}\over{x_2}}f_{j/b}(x_2) \int(dz)}}
\ee
with 
    the lower limit of integration in the variable $z$ given by 
 $z_{min}=4p_{tmin}^2/(sx_1x_2)$. To grasp the energy dependence of
 this scale, one can use a simple toy model, in which the 
 valence quark densities are approximated by $1/\sqrt{x}$ and thus obtain 
\be
\label{qmaxavan}
<q_{max}(s)>\sim {{3}\over{8}} p_{tmin}
 ln{{\sqrt{s}}\over{2p_{tmin}}} 
\ee
for $2p_{tmin}<<\sqrt{s}$. For $p_{tmin}=1.4\ GeV$,  
as in typical eikonal mini-jet models for proton-proton scattering
\cite{BLOCK}, one obtains values of  $<q_{max}(s)>$ which range
from 0.5 to 5 GeV for $\sqrt{s}$ between 10 GeV and 14 TeV respectively.
A more precise
 evaluation of the above quantities depends upon
  the type of parton densities one uses, and will be discussed in a
forthcoming paper.
  \begin{figure}[t]
 \begin{center}
\leavevmode
\mbox{\epsfig{file=starelesnafig2.ps,width=0.7\textwidth,bbllx=30pt,bblly=70pt,bburx=570pt,bbury=750pt,angle=90}}
\end{center}

\caption{ Comparison between the A(b)  distribution function
 from the Bloch-Nordsieck model (full) and the  form factor model (dots).}

\end{figure}

\thispagestyle{plain}
 From the discussion about the large b-behaviour of the function $h(b)$,
   we then expect
  $A_{BN}(b,s)$ to fall at large b more rapidly as the energy increases
   from $\sqrt{s}=10 \ GeV$ 
  into the TeV region. 
In Fig. 2, we compare this behaviour for  the function $A(b)$ with the one
obtained through the
  Fourier transform of the squared e.m. form factor of the proton, eq.(
  \ref{aobpp}).  
 The function
$A(b)$ from the Bloch-Nordsieck model is calculated for $\Lambda=0.1\ GeV$
and 
 values of  $<q_{max}>$ which include those obtainable from
 eq.(\ref{qmaxavan}) in the energy  range
$\sqrt{s}\approx 10\ GeV \div 14\ TeV$.

We notice that, as the energy increases, $A(b) $ from 
 the form factor model remains substantially 
higher at large b 
than in the  Bloch-Nordsieck case.
 As a result, for the same 
$\sigma_{jet}$ the Bloch-Nordsieck model
will give smaller $n(b,s)$ at large b 
than the form factor model and
a softening effect of the total 
eikonal mini-jet cross-sections can be expected. 

\par\vskip 5 mm

In conclusion, we have studied the parameter dependence of the eikonalized
mini-jet model for photonic and hadronic total inelastic cross-section and
found that a large uncertainty arises through the description
of parton distributions in impact parameter space. A model, derived from soft
gluon summation techniques, is described and compared with
expectations from the currently used form factor models
for such disatribution.
Such comparison indicates a distinctly different behaviour in this
 large b-region suggesting a  softening of the rise 
 of the total cross-section in mini-jet models relative 
 to the ones with the hadron  form-factors.

\thispagestyle{plain}

This Work has been partially supported by CICYT, Contract \# AEN 94-00936,
 EEC HCMP CT92-0026, C.S.I.R. (India), grant No. 03 (0745)/94-EMR-II
and US-Department of Energy.

\end{document}